\documentclass{Interspeech}

\interspeechcameraready 
\usepackage{multirow}
\usepackage{cite}
\usepackage{amsmath,amssymb,amsfonts}
\usepackage{algorithmic}
\usepackage{graphicx}
\usepackage{textcomp}
\usepackage{subcaption}
\usepackage{xcolor}
\usepackage{hyperref} 
\usepackage{booktabs} 
\usepackage{multirow}
\title{WAKE: Watermarking Audio with Key Enrichment}

\author[affiliation={1}]{\qquad \qquad \quad Yaoxun}{Xu}
\author[affiliation={2}]{Jianwei}{Yu}
\author[affiliation={2}]{Hangting}{Chen}
\author[affiliation={1,3,*}]{Zhiyong}{Wu}
\author[affiliation={3}]{\\Xixin}{Wu}
\author[affiliation={2}]{Dong}{Yu}
\author[affiliation={2}]{Rongzhi}{Gu}
\author[affiliation={2}]{Yi}{Luo}

\affiliation{Shenzhen International Graduate School}{Tsinghua University}{Shenzhen, China}
\affiliation{}{Tencent AI Lab}{China}
\affiliation{}{The Chinese University of Hong Kong}{Hong Kong SAR, China}

\email{xuyx22@mails.tsinghua.edu.cn, zywu@sz.tsinghua.edu.cn\thanks{* Corresponding author.}}

\usepackage{comment}

\begin{document}

\maketitle

\begin{abstract}
As deep learning advances in audio generation, challenges in audio security and copyright protection highlight the need for robust audio watermarking. Recent neural network-based methods have made progress but still face three main issues: preventing unauthorized access, decoding initial watermarks after multiple embeddings, and embedding varying lengths of watermarks. To address these issues, we propose WAKE, the first key-controllable audio watermark framework. WAKE embeds watermarks using specific keys and recovers them with corresponding keys, enhancing security by making incorrect key decoding impossible. It also resolves the overwriting issue by allowing watermark decoding after multiple embeddings and supports variable-length watermark insertion. WAKE outperforms existing models in both watermarked audio quality and watermark detection accuracy. Code, more results, and demo page: \href{https://thuhcsi.github.io/WAKE}{https://thuhcsi.github.io/WAKE}.
\end{abstract}

\section{Introduction}

The rapid advancement of audio generation models \cite{audio1,audio2,audio3,audio4} has greatly increased both the quantity and quality of audio content. As a result, audio watermarking \cite{wmreview,wmreview1} is becoming essential for preventing misuse and ensuring traceability. This technique embeds a predefined watermark into audio to protect copyright and enable tracking, with the key challenge being to keep the watermark inaudible while ensuring reliable extraction.

Traditional audio watermarking methods \cite{LSB,phase,echo,maha2010dct,dwt,wm1,wm2,wm3} rely on auditory masking \cite{greenwood1961auditory}, but often require manual tuning, leave detectable traces, and have limited capacity. In contrast, deep learning has driven significant advances in audio watermarking. For instance, DeAR \cite{liu2023dear} addresses re-recording attacks, WavMark \cite{chen2023WavMark} employs invertible neural networks, and AudioSeal \cite{roman2024proactive} uses the EnCodec \cite{encodec} framework to achieve effective frame-level watermarking.
%These approaches demonstrate the potential for enhanced robustness and imperceptibility in audio watermarking technology.

Despite substantial progress in audio watermarking research, limitations remain. Firstly, existing methods do not restrict watermark extraction, allowing unauthorized users to freely extract watermarks, especially with public pre-trained models. To enhance security and reduce the risk of external deciphering, users must retrain their own models, increasing costs and complexity.
Secondly, past research focuses on embedding a single watermark, neglecting issues with multiple watermarks. Multiple embeddings can increase capacity and are crucial for traceability during transmission by different entities. However, if a model cannot support multiple watermarks, embedding a second one can prevent tracing the earliest watermark, defeating the purpose of watermarking.
Lastly, current methods only support fixed-length watermarks. To embed varying lengths, users must retrain models for each specific length, increasing costs and limiting scalability.

\begin{figure}[h]
  \centering
  \includegraphics[width=0.42\textwidth]{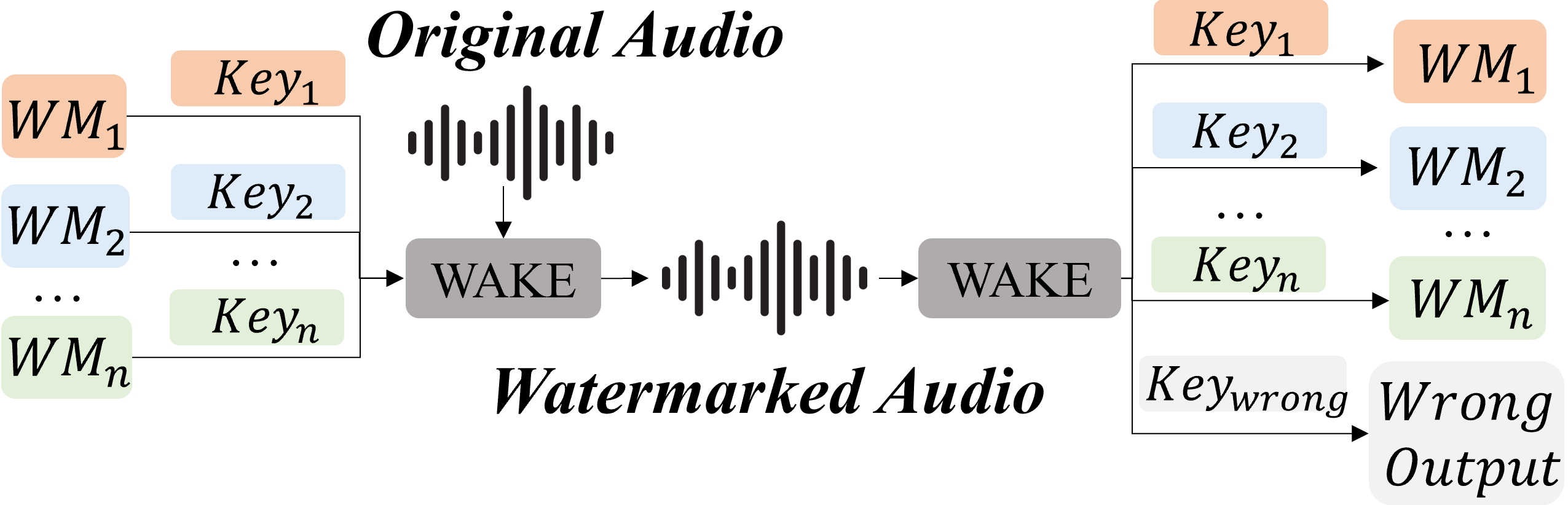}
  \vspace{-2mm}
  \caption{WAKE's overall process: embedding and decoding watermarks (WM) using specific keys.}
  \label{fig1}
  \vspace{-2mm}
\end{figure}
\begin{figure*}[ht]
  \centering
  \includegraphics[width=0.85\textwidth]{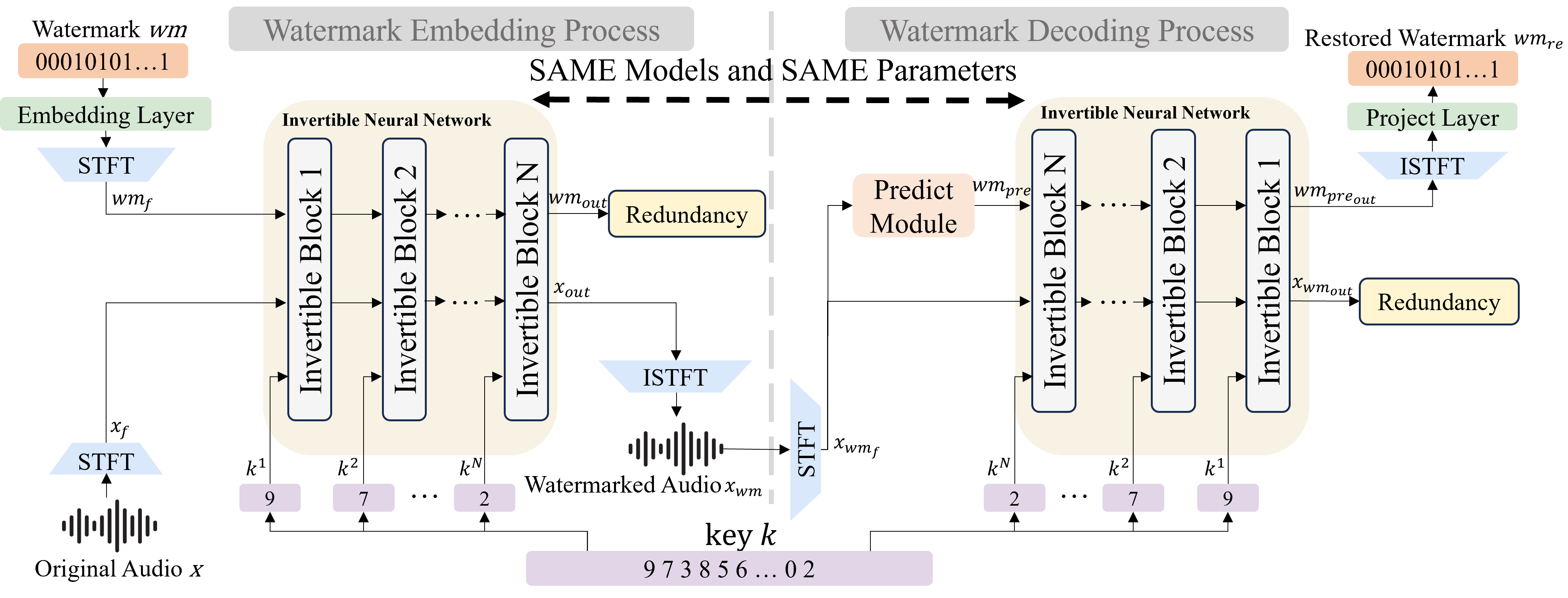}
  \vspace{-2mm}
  \caption{Framework of the proposed WAKE.}
  \label{fig2}
  \vspace{-6mm}
\end{figure*}

To address the aforementioned limitations, we propose WAKE, the first key-controllable audio watermark framework. WAKE embeds watermarks into audio and decodes them using keys, as shown in Figure \ref{fig1}. Using an incorrect key prevents correct watermark decoding. WAKE supports multiple watermark embeddings and corresponding decodings based on keys. Additionally, WAKE enables the embedding of variable-length watermarks for the first time. Our contributions are as follows:
\begin{itemize}
    \item We introduce WAKE, the first key-controllable audio watermarking model, which uniquely embeds and decodes watermarks using specific keys.
    \item WAKE achieves the first multi-watermark embedding and specific watermark extraction in the audio field.
    \item WAKE surpasses current state-of-the-art models in both watermarked audio quality and decoding performance. 
\end{itemize}
\vspace{-1mm}
\section{Methods}
\vspace{-1mm}
%WAKE not only offers personalized key-controllable watermark embedding but also allows for the embedding of multiple different watermarks. Only by using specific keys can we restore the corresponding watermarks. If incorrect keys are used, decaying the correct embedded watermark will be impossible. This approach effectively enhances the security of the watermark and addresses both the capacity issue of audio watermark embedding and the problem of multiple watermark overlays.

\subsection{Model architecture}
As illustrated in Figure \ref{fig2}, WAKE comprises embedding and decoding modules, an invertible neural network (INN), and a Predict Module. Given original audio $x$, watermark $wm$, and key $k$, embedding module begins by converting $x$ to the frequency domain ($x_f$) via STFT. Simultaneously, $wm$ is embedded and transformed to $wm_f$. The INN, guided by $k$, processes $x_f$ and $wm_f$ to produce $x_{out}$ and $wm_{out}$, where $wm_{out}$ serves as redundancy. $x_{out}$ is then converted back to the time domain with ISTFT, yielding the watermarked audio $x_{wm}$.
For decoding, $x_{wm}$ is transformed to $x_{wm_f}$ via STFT. The Predict Module extracts the initial watermark feature $wm_{pre}$ from $x_{wm_f}$. The INN then reverses the process using $x_{wm_f}$, $wm_{pre}$, and $k$ to recover $x_{wm_{out}}$ (redundancy) and $wm_{pre_{out}}$. Finally, ISTFT and a mapping layer reconstruct the restored watermark $wm_{re}$ from $wm_{pre_{out}}$.

\subsubsection{Invertible neural network}
\begin{figure}[!h]
\centering
\includegraphics[width=0.5\textwidth]{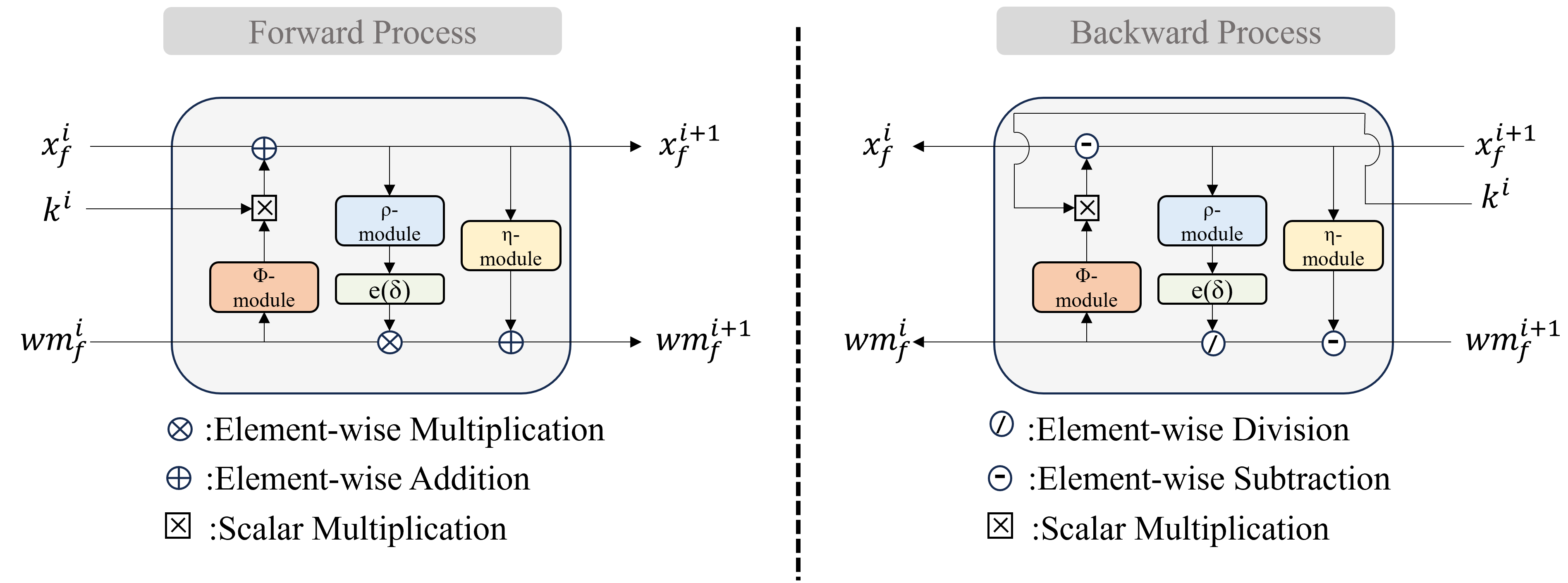}
\caption{The structure of INN block $i$.}
\label{fig3}
\vspace{-3mm}
\end{figure}

Inspired by audio \cite{chen2023WavMark}, image \cite{imagesteg,imagesteg1,fang2023flow}, and video \cite{videosteg}, we use an INN as the core of WAKE due to its high reusability and computational efficiency. The INN comprises $N$ invertible blocks, with the structure of block $i$ shown in Figure \ref{fig3}. The key length $N$, matches the number of invertible blocks in the INN. Each bit $k^i$ of the key corresponds to invertible block $i$.

The forward process of invertible block $i$ is as follows:
\vspace{-0.05mm}
\begin{equation}
\begin{aligned}
x_f^{i+1} &= x_f^i + \phi(wm_f^i) \cdot k^i, \\
wm_f^{i+1} &= wm_f^i \odot \exp(\alpha(\rho(x_f^{i+1}))) + \eta(x_f^{i+1}).
\end{aligned}
\end{equation}
Here, $\alpha$ represents the clamp function, $\rho$, $\eta$, and $\phi$ represent transformation, and $\odot$ denotes element-wise multiplication.

\begin{table*}[!h]
\centering
\caption{Experimental results on single and double watermarking. $BER_i^j$ represents the BER between the $j^{th}$ embedded watermark and the watermark decoded with the $i^{th}$ key by WAKE or directly decoded by AudioSeal and WavMark. $i$=3 denotes using a distinct key, different from the first two keys, which indicates the case of decoding with the incorrect key. Values after $\pm$ are standard deviations.}
\vspace{-2mm}
\label{table1}
\scalebox{0.8}{
% \usepackage{multirow}

% \usepackage{multirow}

\begin{tabular}{c|c|ccccccc}
\toprule[2pt]
\begin{tabular}[c]{@{}c@{}}Watermark\\ Scenario\end{tabular}                   & Model     & bit & $SNR$   {\small$\uparrow$}                                            & $PESQ$      {\small$\uparrow$}                                        & $BER_1^1$  {\small$\downarrow$}                                       & $BER_3^1$                                & $BER_2^2$  {\small$\downarrow$}                                        & $BER_3^2$                                \\ \hline
\multirow{3}{*}{\begin{tabular}[c]{@{}c@{}}Single\\ Watermark\end{tabular}} & AudioSeal & 16  & 24.686{\small$\pm$0.120}          & 4.316{\small$\pm$0.010}           & 1.910{\small$\pm$0.108}          & 1.910{\small$\pm$0.108}   & -                                                & -                                        \\
                                                                            & WavMark   & 32  & 38.594{\small$\pm$0.130}          & 4.256{\small$\pm$0.018}          & 1.091{\small$\pm$0.139}          & 1.090{\small$\pm$0.139}  & -                                                & -                                        \\
                                                                            & WAKE      & 32  & \textbf{41.237{\small$\pm$0.121}} & \textbf{4.396{\small$\pm$0.016}} & \textbf{0.123{\small$\pm$0.002}} & 50.090{\small$\pm$0.075} & -                                                & -                                        \\ \hline
\multirow{3}{*}{\begin{tabular}[c]{@{}c@{}}Double\\ Watermark\end{tabular}} & AudioSeal & 32  & 22.132{\small$\pm$0.130}          & 4.200{\small$\pm$0.006}           & 46.750{\small$\pm$0.403}         & 46.750{\small$\pm$0.403} & 3.790{\small$\pm$0.188}          & 3.790{\small$\pm$0.188}   \\
                                                                            & WavMark   & 64  & 35.817{\small$\pm$0.110}          & 4.181{\small$\pm$0.009}          & 48.370{\small$\pm$0.245}         & 48.370{\small$\pm$0.245} & \textbf{1.910{\small$\pm$0.112} } & 1.910{\small$\pm$0.112}  \\
                                                                            & WAKE      & 64  & \textbf{38.884{\small$\pm$0.130}} & \textbf{4.318{\small$\pm$0.017}} & \textbf{1.252{\small$\pm$0.091}} & 42.460{\small$\pm$0.201} & 2.709{\small$\pm$0.126}          & 41.440{\small$\pm$0.303} \\ \bottomrule[2pt]
\end{tabular}
}
\vspace{-2mm}

\end{table*}

\subsubsection{Predict module}

In INN-based systems, fully restoring the watermark requires back-propagating the redundancy from the forward process. However, this redundancy isn’t directly accessible during decoding. Most INN-based studies \cite{chen2023WavMark,kingma2018glow,fang2023flow,inn1,inn2} use Gaussian-distributed redundancies and random Gaussian sampling in the backward process. We argue that, since the watermark interacts with the audio during embedding, the redundancy should be audio-dependent rather than random. As a result, random sampling complicates watermark recovery.
\begin{figure}[!h]
\centering
\includegraphics[width=0.35\textwidth]{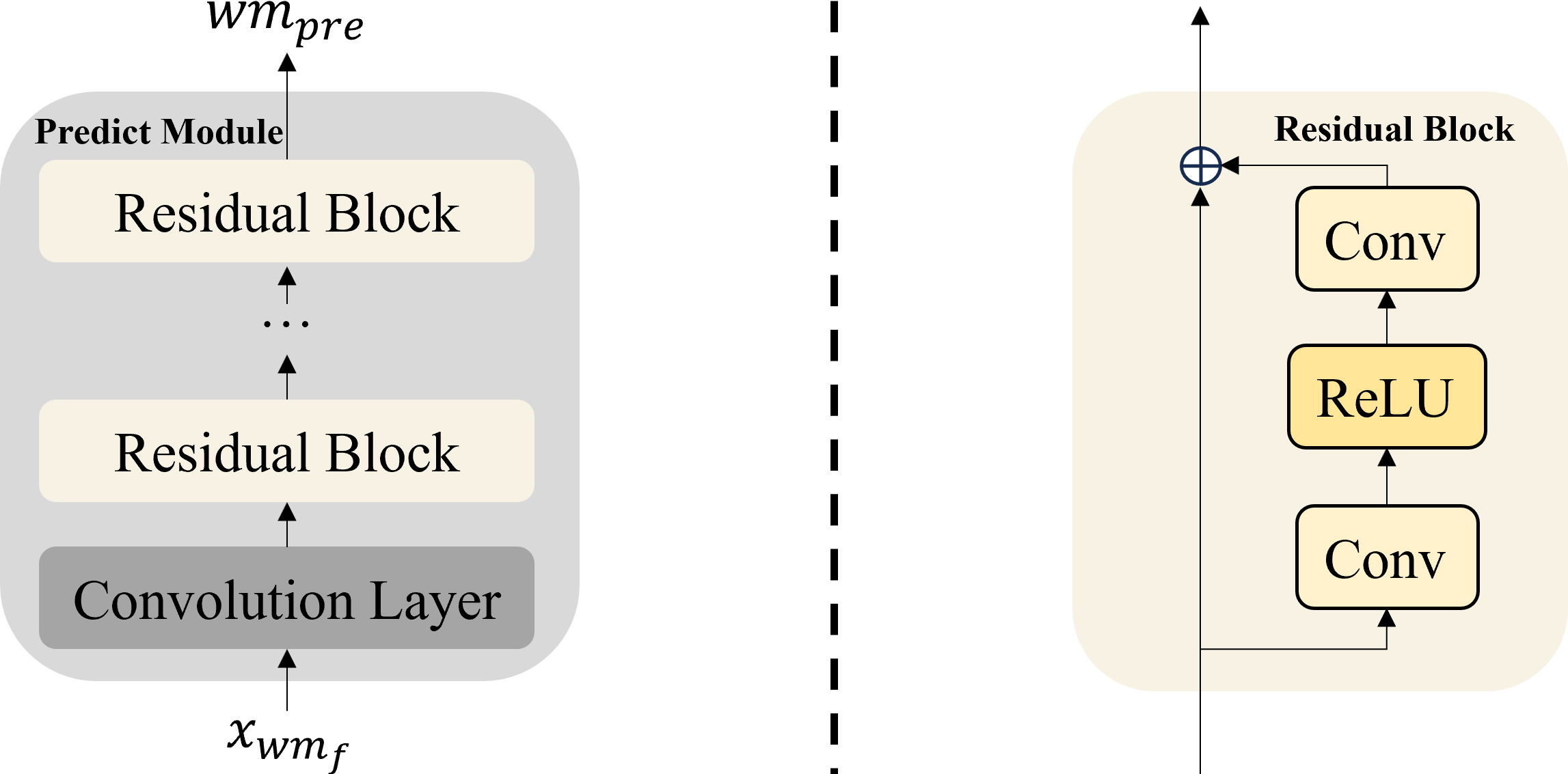}
\caption{Predict Module and Residual Block Structure.}
\vspace{-3mm}
\label{fig4}
\end{figure}

Based on this, we design the Predict Module (Figure \ref{fig4}), generating ${wm}_{pre}$ for INN decoding. Since ${wm}_{pre}$ is related to audio features, it restores the watermark more effectively.

\subsection{Audio editing operation}
To improve WAKE's robustness in decoding watermarks affected by audio editing, we apply common editing operations such as up-down sampling (UD), random noise (RN), pink noise (PN), low-pass filters (LF), high-pass filters (HF), band-pass filters (BF), boost audio (BA), duck audio (DA), and shush attacks (SA). During training, we randomly select one of these operations and input the processed audio into the decoding process, enhancing WAKE's adaptability to audio editing.

\subsection{Training process}
\subsubsection{Loss function}
WAKE aims to embed watermarks into audio so that the watermarked and original audio are perceptually indistinguishable, and to decode the embedded watermarks accurately. To achieve this, we design perceptual and accuracy constraints.

\noindent\textbf{Perceptual constraints} To keep the watermarked audio indistinguishable from the original, we apply an L2 loss in the time domain, a multi-scale Mel-spectrogram loss \cite{encodec} in the frequency domain, and adversarial training with a discriminator. However, as shown in our experiments (Sec 4.3), these constraints alone led to watermark artifacts at audio edges, causing booming sounds and excess high-frequency energy. To address this, we introduced stronger BroadWeight constraints on segment boundaries, based on the multi-scale Mel-spectrogram loss, to improve audio consistency at the edges. The perceptual constraint is defined as:
%\small{
\begin{equation}
\begin{aligned}
\mathcal{L}_p({x},{x}_{wm}) &= w_{p1}||{x} - {x}_{wm}||_2 + w_{p2} \log(1 - D({x}_{wm})) +\\
& w_{p3} \sum_{i \in e} \left( ||S_i({x} \odot W) - S_i({x}_{wm} \odot W)||_1 + \right.\\
& \left. ||S_i({x} \odot W) - S_i({x}_{wm} \odot W)||_2 \right).
\end{aligned}
\end{equation}
%}
In this equation, ${x}$ is the original audio, ${x}_{wm}$ is the watermarked audio, $||\cdot||_1$ is the L1 loss, $||\cdot||_2$ is the L2 loss, $S_i$ is a 64-bin Mel-spectrogram using a normalized STFT with a window size of $2^i$ and hop length of $2^i/4$, $e = \{5, \ldots, 11\}$ is the set of scales, $W$ represents the BroadWeight constraints, and $D(\cdot)$ is the discriminator for adversarial training with WAKE. $w_{p1}$, $w_{p2}$, and $w_{p3}$ are the weights for the respective terms.

\noindent\textbf{Accuracy constraints}
To ensure accurate watermark decoding, we use binary cross-entropy loss (BCELoss) to make the decoded watermark $wm_{re}$, obtained with the correct key, closer to the embedded watermark $wm$. We also aim for WAKE to produce inconsistent watermarks with an incorrect key. During training, a random incorrect key $k_{\text{wrong}}$ generates a watermark $wm_{\text{wrong}}$, which we expect to differ from $wm$ without limiting their dissimilarity. The accuracy constraints are as follows:
\begin{equation}
\begin{aligned}
\mathcal{L}_a(&wm, wm_{re}) = \textit{BCELoss}(wm, wm_{re}) + \\
& w_{l1} * \max(0, w_{l2} - \textit{BCELoss}(wm, wm_{wrong})).
\end{aligned}
\end{equation}
Here, $w_{l1}$ controls the weight of decoding with an incorrect key, while $w_{l2}$ is the set threshold.

\subsubsection{Training strategy}

To enhance WAKE's ability to embed and decode multiple watermarks with specific keys, in each training step, we randomly choose one of the following methods to ensure WAKE handles both single and multiple watermarks.

1. Generate a watermark $wm$ and a key $k$, then embed $wm$ into the original audio ${x}$ using $k$ to obtain watermarked audio ${x}_{wm}$. After audio editing, this becomes ${x}_e$. Decode ${x}_e$ using $k$ to retrieve the watermark $wm_{re}$. The training loss is:
%{\small{
\begin{equation}
\mathcal{L}_{train} = w_{t1}\mathcal{L}_p({x},{x}_{wm}) + w_{t2}\mathcal{L}_a(wm, wm_{re}),
\end{equation}
%}}
where $w_{t1}$ and $w_{t2}$ are the weights for perceptual and accuracy constraints, respectively.

2. Generate two different watermarks $wm_1$ and $wm_2$ with keys $k_1$ and $k_2$. Embed $wm_1$ into ${x}$ using $k_1$ to get ${x}_{wm1}$, then embed $wm_2$ into ${x}_{wm1}$ using $k_2$ to get ${x}_{wm2}$. After audio editing, this becomes ${x}_e$. Decode ${x}_e$ using both $k_1$ and $k_2$ to retrieve $wm_{re1}$ and $wm_{re2}$. The training loss is:
%{\small{
\begin{equation}
\begin{aligned}
\mathcal{L}_{train} &= w_{t1}(\mathcal{L}_p({x},{x}_{wm1}) + \mathcal{L}_p({x},{x}_{wm2})) + \\
& w_{t2}(\mathcal{L}_a(wm_1, wm_{re1}) + \mathcal{L}_a(wm_2, wm_{re2})).
\end{aligned}
\end{equation}

\begin{table*}[h]
\vspace{-1mm}
\caption{Comparison of BER for decoded watermarks subjected to various audio editing operations. WAKE-$i$-$j$ denotes decoding with the $j^{th}$ key after embedding $i$ different watermarks. ``NA" indicates that no audio editing operation is applied.}
\vspace{-2mm}
\label{table2}
\centering
\scalebox{0.8}{
\begin{tabular}{ccccccccccc}
\toprule[2pt]
   Model       & NA             & UD             & RN             & PN             & LF             & HF             & BF             & BA             & DA             & SA             \\ \hline
AudioSeal & 1.911          & 1.911          & 2.200          & 1.929          & 1.924          & 2.104          & 2.151          & 1.984          & 1.904          & 49.806         \\
WavMark   & 1.003          & 0.994          & 5.142          & 1.133          & 2.053          & 1.027          & 1.049          & 1.008          & 1.099          & 0.990          \\
WAKE-1-1  & \textbf{0.123} & \textbf{0.173} & \textbf{1.028} & \textbf{0.130} & \textbf{1.765} & \textbf{0.131} & \textbf{0.129} & \textbf{0.127} & \textbf{0.125} & \textbf{0.127} \\
WAKE-2-1  & 1.252          & 1.266          & 2.527          & 1.259          & 3.261          & 1.266          & 1.258          & 1.262          & 1.274          & 1.233          \\
WAKE-2-2  & 2.709          & 2.817          & 5.016          & 2.762          & 6.653          & 2.721          & 2.717          & 2.708          & 2.731          & 2.681          \\ \bottomrule[2pt]
\end{tabular}
}
\vspace{-2mm}
\end{table*}

\begin{table*}[h]
\centering
\caption{Impact of different perceptual constraints on watermarking performance.}
\vspace{-3mm}
\label{table3}
\scalebox{0.8}{
% \usepackage{multirow}
\begin{tabular}{c|c|ccc|cccc}
\toprule[2pt]
         \multirow{2}{*}{Perceptual Constraints}                & \multirow{2}{*}{ID} & \multicolumn{3}{c|}{Single Watermark} & \multicolumn{4}{c}{Double Watermark}   \\ \cline{3-9} 
                         &                     & $SNR$  {\small$\uparrow$}         & $PESQ$  {\small$\uparrow$}      & $BER_1^1$  {\small$\downarrow$}        & $SNR$  {\small$\uparrow$}   & $PESQ$ {\small$\uparrow$}  & $BER_1^1$ {\small$\downarrow$}   & $BER_2^2$ {\small$\downarrow$}  \\ \hline
L2 Loss                  & \#1                 & 35.113       & 4.023      & 0.14      & 32.832 & 3.972 & 1.42      & 3.88      \\
\#1+Multi-Scale Mel Loss & \#2                 & 39.344       & 4.231      & \textbf{0.08}      & 37.072 & 4.232 & \textbf{1.02}      & \textbf{2.56}     \\
\#2+BroadWeight (WAKE)   & \#3                 & \textbf{41.103}       & \textbf{4.388}      & 0.12      & \textbf{38.893} & \textbf{4.311} & 1.26      & 2.71      \\ \bottomrule[2pt]

\end{tabular}
}
\vspace{-3mm}

\end{table*}

\section{Experiment setting}
\label{experimentset}

\noindent\textbf{Dataset} WAKE uses 3,529.7 hours of different types of training data including LibriSpeech \cite{panayotov2015librispeech}, Common Voice \cite{ardila2019common}, Audio Set \cite{audioset}, and Free Music Archive \cite{defferrard2016fma}. We selected 500 samples from each dataset for validation and testing.

\noindent \textbf{Basic settings} 
During training, the audio sampling rate is set to 16,000 Hz for a duration of 1 second. Each embedded watermark and key length are 32 and 8, respectively. WAKE's INN consists of 8 invertible blocks, matching the key length, and each block includes 5 layers of 2D CNNs with dense connections\cite{chen2023WavMark,inn1}. We use a window length of 1,000 and a shift of 400 for STFT. The Predict Module contains 8 residual blocks.

\noindent \textbf{Training weights} We set $W$ for the first and last 3\% of the time to 10, while the others are set to 1. $w_{p1}$, $w_{p2}$, $w_{p3}$, $w_{t1}$, $w_{t2}$, $w_{l1}$, $w_{l2}$ are 1, 1, 5, 10, 10, 1000, and 0.01 respectively.

% \noindent \textbf{Training equipment} WAKE is trained on 8 32GB V100 GPUs with a batch size of 4, using the Adam optimizer for 600k steps and a learning rate of 1e-5.

\noindent \textbf{Baselines} To evaluate WAKE, we used the current best WavMark (32 bps) and AudioSeal (16 bps) as baselines.  WAKE can reach 32 * $n$ bps, where $n$ is the embedding times.

\noindent \textbf{Evaluation metrics} We assess the watermarked audio's imperceptibility using Perceptual Evaluation of Speech Quality (PESQ) \cite{pesq} and Signal-to-Noise Ratio (SNR), and evaluate decoding accuracy with Bit Error Rate (BER).

\vspace{-1mm}
\section{Results}
\label{result}
This section presents a detailed analysis of WAKE's performance in generating and decoding watermarked audio with specific keys, based on comprehensive experiments.\footnote{In this section, we primarily focus on discussing single and double watermark embedding. A comparison of experiments with multiple watermark embedding and more results can be found in the \href{https://xuyaoxun.github.io/WAKE_demo/}{\textbf{demo page}}.}

Following training, we use 1-second audio clips for watermark embedding, conducting each test five times to minimize randomness. A 32-bit watermark is randomly generated for all models to ensure fairness. For AudioSeal, only the first 16 bits are used. An 8-bit key is randomly generated for each trial to control WAKE's embedding. In multiple watermark experiments, we ensure unique watermarks and keys in each trial.
\vspace{-2mm}
\subsection{Main results on single and double watermark scenarios}
\vspace{-1mm}
This experiment compares WAKE's performance with baselines for single and double watermark scenarios. Since AudioSeal and WavMark do not support key-based decoding, we only compare directly decoded watermarks, as shown in Table \ref{table1}.

In the single watermark scenarios, WAKE outperforms AudioSeal and WavMark in audio perception (SNR and PESQ) and decoding accuracy. With the correct key, WAKE achieves a low BER of 0.12\%. With an incorrect key, the BER is 50.09\%, approximating random guessing, demonstrating WAKE's effectiveness in preventing correct decoding with incorrect keys.

In the double watermark scenario, increased embedding capacity affects audio quality. However, WAKE surpasses baselines in SNR and PESQ, matching baselines in the single-watermark scenario. WAKE effectively decodes the first and second watermarks using their respective keys, with BERs of 1.25\% and 2.71\%.
In contrast, AudioSeal and WavMark have BERs of 46.75\%/3.79\% and 48.37\%/1.91\% for the first and second watermarks, respectively, indicating they can only decode the second watermark and fail to decode the first. This vulnerability is problematic, as an audio watermarking system where subsequent watermarks overwrite previous ones is highly susceptible to attacks, posing significant security risks.
Similar to the single watermark scenario, if an incorrect key is used, WAKE cannot decode either watermark correctly, with error rates approaching 50\%. This shows WAKE's decoding is strongly key-dependent, enhancing system security.
Furthermore, this highlights the potential of embedding watermarks of varying lengths by incorporating multiple watermarks and decoding them with their respective keys, thereby enhancing the capacity and scalability of audio watermarking.
\vspace{-2mm}
\subsection{Decoding performance with audio editing operations}
\vspace{-1mm}
\label{attack}

To evaluate robustness against audio editing, we test AudioSeal and WavMark with single watermark embedding, while WAKE is assessed under three scenarios: (1) decoding with the correct key after embedding a single watermark, (2) decoding with the first key after embedding two watermarks, and (3) decoding with the second key after embedding two watermarks. Results are presented in Table \ref{table2}.

WAKE achieves the strongest resistance to audio editing with a single watermark. Although decoding becomes more challenging with multiple watermarks, WAKE still maintains robust performance under various editing operations.

\vspace{-2mm}
\subsection{The impact of different perceptual constraints}
\vspace{-1mm}
% \usepackage{multirow}

This study investigates the effects of various constraints on WAKE's perceptual quality, as displayed in Table \ref{table3}.

% \usepackage{multirow}

The experiment shows that using only L2 loss results in lower SNR and PESQ. Integrating multi-scale Mel loss significantly boosts SNR and PESQ from 35.113/4.023 to 39.344/4.231 for a single watermark, with noticeable improvements for double watermarks as well. Introducing BroadWeight constraints further improves SNR and PESQ values, but also increases BER, likely due to the model's focus on audio segment edges complicating watermark decoding.

\begin{figure}[ht]
\vspace{-3mm}
    \centering
    \subfloat[Original Audio]{%
        \includegraphics[width=0.13\textwidth]{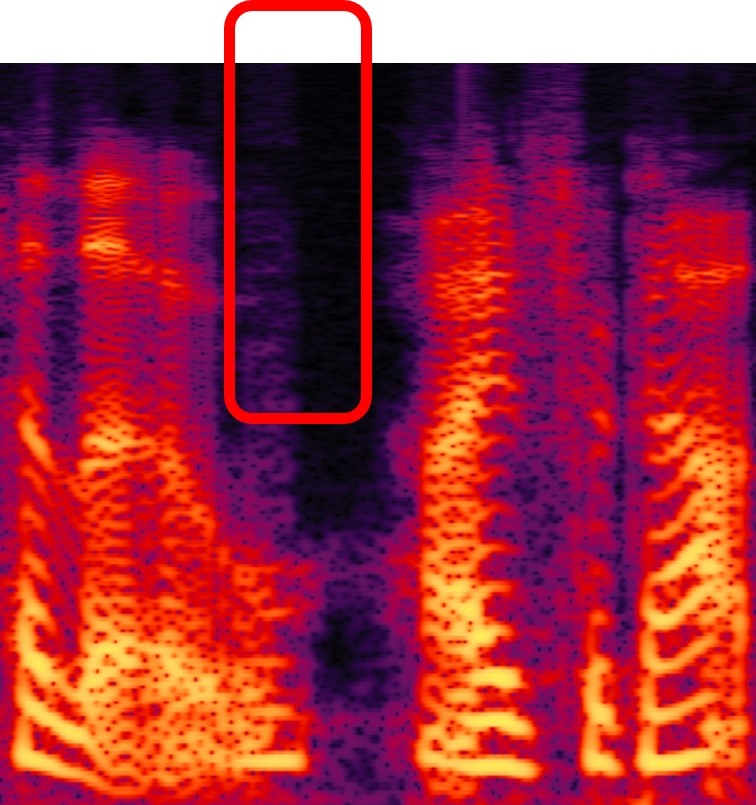}%
        \label{fig:original}
    }\hspace{8mm}
    \subfloat[\#1 Constraints]{%
        \includegraphics[width=0.13\textwidth]{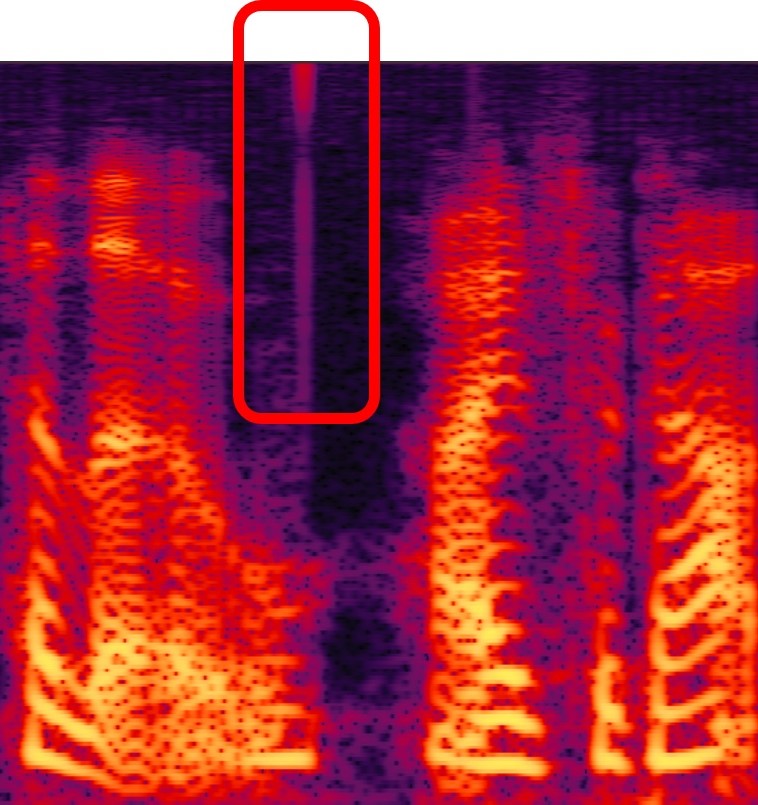}%
        \label{52}
    }\\

    \subfloat[\#2 Constraints]{%
        \includegraphics[width=0.13\textwidth]{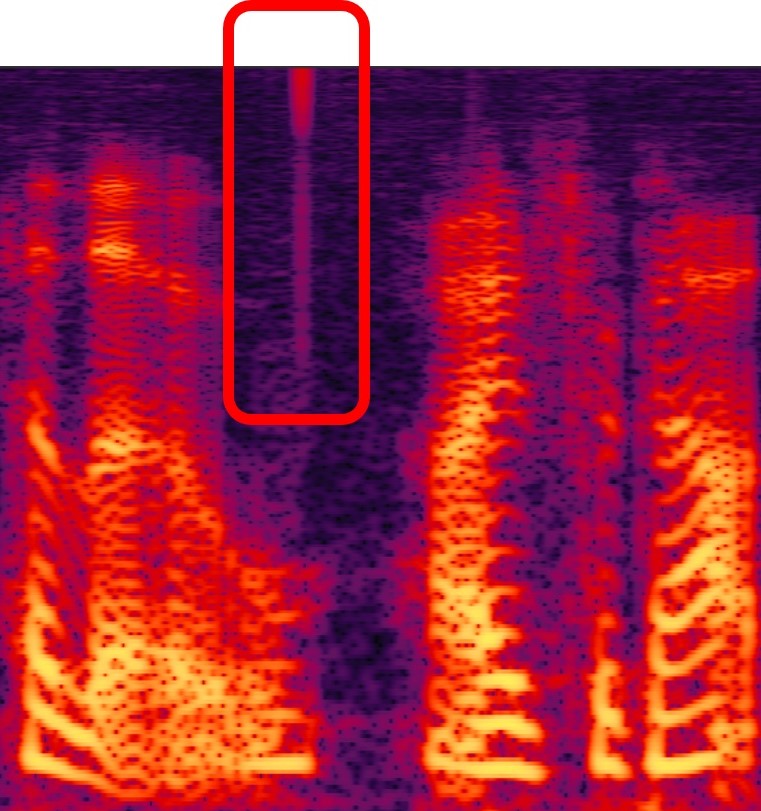}%
        \label{53}
    }\hspace{8mm}
    \subfloat[\#3 Constraints]{%
        \includegraphics[width=0.13\textwidth]{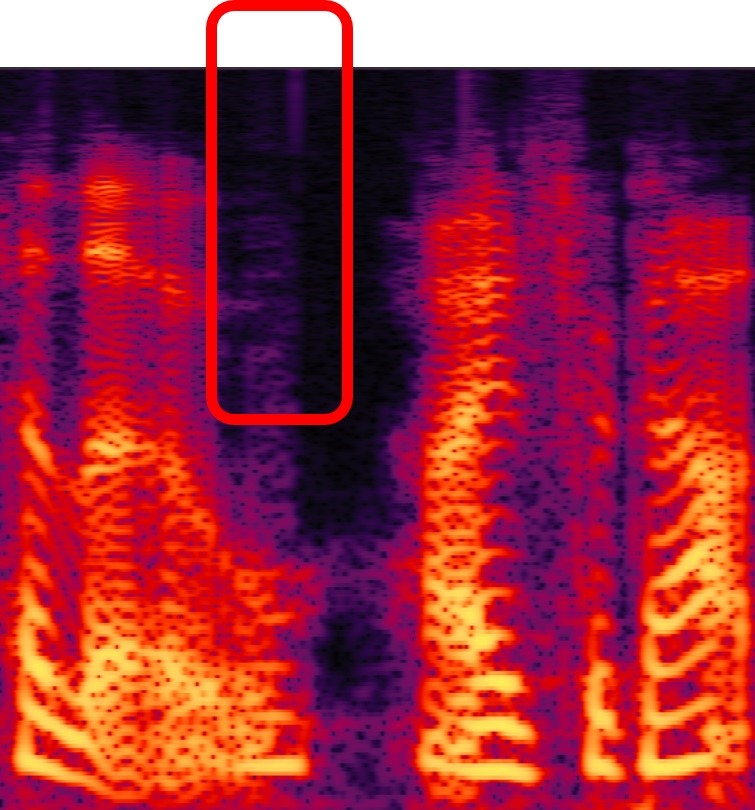}%
        \label{54}
    }
    \vspace{-3mm}
    \caption{Comparison of different perceptual constraints.}
    \label{fig5}
    \vspace{-3mm}
\end{figure}

SNR and PESQ do not directly reflect human perceptual evaluation. For example, a slight bursting sound is observed at the watermarked audio edges by \#1 and \#2. Comparing spectrograms of the same audio with the same watermark under three constraints (Figure \ref{fig5}), we find that without BroadWeight constraints, WAKE shows significant energy stacking in the high-frequency region at audio edges (red box in Figure \ref{52} and \ref{53}), causing a noticeable bursting sound. With BroadWeight constraints, the energy distribution (Figure \ref{54}) at the audio edges resembles the original audio, eliminating the bursting sound.

% \begin{figure}[h]
%   \centering
%   \includegraphics[width=1.0\textwidth]{fig/fig5.png}
%   \caption{The figure below shows the spectrograms of the original audio, the audio with the watermark embedded using L2 loss, the audio with the watermark embedded using multi-scale Mel-spectrum loss, and the audio with the watermark embedded using edge enhancement loss.}
%   \label{fig5}
% \end{figure}
\vspace{-2mm}
% \subsection{Balance between audio quality and decoding accuracy}
% \vspace{-1mm}
% % Please add the following required packages to your document preamble:
% % \usepackage{multirow}
% Audio quality and accurate watermark decoding are both crucial for audio watermarking models. To balance these aspects, we adjust the accuracy constraint weight $w_{t2}$ in the training loss while keeping the perceptual constraint weight $w_{t1}$ fixed. Results are shown in Table \ref{table4}.

% The experiments show that as $w_{t2}$ increases, both SNR and PESQ decrease, reducing audio quality. Conversely, the BER for single and double watermarked scenarios drops respectively, indicating improved decoding capability. Pursuing better decoding performance leads to lower sound quality, and vice versa. Therefore, we choose $w_{t2}$=10 for the optimal balance between audio quality and decoding capability.

\subsection{Ablation study of the Predict Module}

In this study, we examine the Predict Module's impact on WAKE, with results shown in Table \ref{table5}.

\begin{table}[h]
\vspace{-3mm}
\centering
\caption{Ablation study on the effect of the Predict Module (PM) on model performance.}
\vspace{-3mm}
\scalebox{0.76}{
\begin{tabular}{cc@{\hspace{1.2mm}}c@{\hspace{1.2mm}}c|c@{\hspace{1.2mm}}c@{\hspace{1.2mm}}c@{\hspace{1.2mm}}c}
\toprule[2pt]
                   & \multicolumn{3}{c|}{Single Watermark} & \multicolumn{4}{c}{Double Watermark}    \\ \cline{2-8} 
                   & $SNR$   {\small$\uparrow$}       & $PESQ$  {\small$\uparrow$}     & $BER_1^1$  {\small$\downarrow$}       & $SNR$  {\small$\uparrow$}   & $PESQ$ {\small$\uparrow$}  & $BER_1^1$ {\small$\downarrow$}  & $BER_2^2$ {\small$\downarrow$}  \\ \hline
w/o PM & 36.930      & 4.223     & 5.10     & 34.250  & 4.102 & 8.92     & 13.24     \\
w/ PM   & 41.227      & 4.392     & 0.13     & 38.442 & 4.321 & 1.25     & 2.72    \\
\bottomrule[2pt]
\end{tabular}
}
\vspace{-2mm}

\label{table5}
\end{table}

Results show that removing the Predict Module significantly reduces WAKE’s decoding performance, resulting in a marked increase in BER for both single and double watermark scenarios. Using the Predict Module’s predicted features, which are closely related to the watermarked audio, instead of random sampling, greatly improves both decoding accuracy and audio quality. This is likely due to WAKE’s invertible architecture, where effective constraints during decoding provide better guidance for encoding. In contrast, random Gaussian sampling during training makes convergence more difficult and weakens watermark encoding. Overall, the Predict Module enables more reliable watermark recovery and higher audio quality.

\vspace{-1mm}
\section{Conclusion}
This study presents WAKE, the first key-controllable audio watermarking model that uses specific keys for both embedding and decoding watermarks. Using an incorrect key prevents the correct watermark from being decoded, enhancing security. Furthermore, WAKE effectively solves the problem of previously embedded watermarks being undetectable after multiple watermark insertions by using different keys for each watermark embedding and extracting them with the corresponding key during extraction, significantly enhancing the embedding capacity. For the first time, it allows for the embedding of watermarks of varying lengths, improving usability and scalability. Additionally, WAKE outperforms existing state-of-the-art audio watermarking models in both audio quality and watermark decoding accuracy.
\vspace{-2mm}
\section{Acknowledgements}
This work is supported by National Natural Science Foundation of China (62076144) and Shenzhen Science and Technology Program (JCYJ20220818101014030).
\bibliographystyle{IEEEtran}
\bibliography{mybib}

\begin{thebibliography}{10}
\providecommand{\url}[1]{#1}
\csname url@samestyle\endcsname
\providecommand{\newblock}{\relax}
\providecommand{\bibinfo}[2]{#2}
\providecommand{\BIBentrySTDinterwordspacing}{\spaceskip=0pt\relax}
\providecommand{\BIBentryALTinterwordstretchfactor}{4}
\providecommand{\BIBentryALTinterwordspacing}{\spaceskip=\fontdimen2\font plus
\BIBentryALTinterwordstretchfactor\fontdimen3\font minus \fontdimen4\font\relax}
\providecommand{\BIBforeignlanguage}[2]{{%
\expandafter\ifx\csname l@#1\endcsname\relax
\typeout{** WARNING: IEEEtran.bst: No hyphenation pattern has been}%
\typeout{** loaded for the language `#1'. Using the pattern for}%
\typeout{** the default language instead.}%
\else
\language=\csname l@#1\endcsname
\fi
#2}}
\providecommand{\BIBdecl}{\relax}
\BIBdecl

\bibitem{audio1}
M.~Le, A.~Vyas, B.~Shi, B.~Karrer, L.~Sari, R.~Moritz, M.~Williamson, V.~Manohar, Y.~Adi, J.~Mahadeokar \emph{et~al.}, ``Voicebox: Text-guided multilingual universal speech generation at scale,'' \emph{Advances in neural information processing systems}, vol.~36, 2024.

\bibitem{audio2}
Z.~Ju, Y.~Wang, K.~Shen, X.~Tan, D.~Xin, D.~Yang, Y.~Liu, Y.~Leng, K.~Song, S.~Tang \emph{et~al.}, ``Naturalspeech 3: Zero-shot speech synthesis with factorized codec and diffusion models,'' \emph{arXiv preprint arXiv:2403.03100}, 2024.

\bibitem{audio3}
C.~Du, Y.~Guo, F.~Shen, Z.~Liu, Z.~Liang, X.~Chen, S.~Wang, H.~Zhang, and K.~Yu, ``Unicats: A unified context-aware text-to-speech framework with contextual vq-diffusion and vocoding,'' in \emph{Proceedings of the AAAI Conference on Artificial Intelligence}, vol.~38, 2024, pp. 17\,924--17\,932.

\bibitem{audio4}
Z.~Evans, J.~D. Parker, C.~Carr, Z.~Zukowski, J.~Taylor, and J.~Pons, ``Long-form music generation with latent diffusion,'' \emph{arXiv preprint arXiv:2404.10301}, 2024.

\bibitem{wmreview}
G.~Hua, J.~Huang, Y.~Q. Shi, J.~Goh, and V.~L. Thing, ``Twenty years of digital audio watermarking—a comprehensive review,'' \emph{Signal processing}, vol. 128, pp. 222--242, 2016.

\bibitem{wmreview1}
E.~Salah, Z.~Narima, A.~Khaldi, and K.~M. Redouane, ``Survey of imperceptible and robust digital audio watermarking systems,'' \emph{Multimedia Tools and Applications}, pp. 1--47, 2024.

\bibitem{LSB}
H.~A. Nassrullah, W.~N. Flayyih, and M.~A. Nasrullah, ``Enhancement of lsb audio steganography based on carrier and message characteristics.'' \emph{J. Inf. Hiding Multim. Signal Process.}, vol.~11, no.~3, pp. 126--137, 2020.

\bibitem{phase}
F.~Djebbar, B.~Ayad, H.~Hamam, and K.~Abed-Meraim, ``A view on latest audio steganography techniques,'' in \emph{2011 International Conference on Innovations in Information Technology}.\hskip 1em plus 0.5em minus 0.4em\relax IEEE, 2011, pp. 409--414.

\bibitem{echo}
B.-S. Ko, R.~Nishimura, and Y.~Suzuki, ``Time-spread echo method for digital audio watermarking,'' \emph{IEEE Transactions on Multimedia}, vol.~7, no.~2, pp. 212--221, 2005.

\bibitem{maha2010dct}
C.~Maha, E.~Maher, K.~Mohamed, and B.~A. Chokri, ``Dct based blind audio watermarking scheme,'' in \emph{2010 International conference on signal processing and multimedia applications (SIGMAP)}.\hskip 1em plus 0.5em minus 0.4em\relax IEEE, 2010, pp. 139--144.

\bibitem{dwt}
N.~K. Kalantari, S.~M. Ahadi, and A.~Kashi, ``A robust audio watermarking scheme using mean quantization in the wavelet transform domain,'' in \emph{2007 IEEE International Symposium on Signal Processing and Information Technology}.\hskip 1em plus 0.5em minus 0.4em\relax IEEE, 2007, pp. 198--201.

\bibitem{wm1}
Y.~Lin, W.~H. Abdulla \emph{et~al.}, ``Audio watermark,'' \emph{Audio Watermark A Comprehensive Foundation Using MATLAB}, 2015.

\bibitem{wm2}
G.~Hua, J.~Huang, Y.~Q. Shi, J.~Goh, and V.~L. Thing, ``Twenty years of digital audio watermarking—a comprehensive review,'' \emph{Signal processing}, vol. 128, pp. 222--242, 2016.

\bibitem{wm3}
P.~Bassia, I.~Pitas, and N.~Nikolaidis, ``Robust audio watermarking in the time domain,'' \emph{IEEE Transactions on multimedia}, vol.~3, no.~2, pp. 232--241, 2001.

\bibitem{greenwood1961auditory}
D.~D. Greenwood, ``Auditory masking and the critical band,'' \emph{The journal of the acoustical society of America}, vol.~33, no.~4, pp. 484--502, 1961.

\bibitem{liu2023dear}
C.~Liu, J.~Zhang, H.~Fang, Z.~Ma, W.~Zhang, and N.~Yu, ``Dear: A deep-learning-based audio re-recording resilient watermarking,'' in \emph{Proceedings of the AAAI Conference on Artificial Intelligence}, vol.~37, 2023, pp. 13\,201--13\,209.

\bibitem{chen2023WavMark}
G.~Chen, Y.~Wu, S.~Liu, T.~Liu, X.~Du, and F.~Wei, ``Wavmark: Watermarking for audio generation,'' \emph{arXiv preprint arXiv:2308.12770}, 2023.

\bibitem{roman2024proactive}
R.~S. Roman, P.~Fernandez, A.~D{\'e}fossez, T.~Furon, T.~Tran, and H.~Elsahar, ``Proactive detection of voice cloning with localized watermarking,'' \emph{arXiv preprint arXiv:2401.17264}, 2024.

\bibitem{encodec}
A.~D{\'e}fossez, J.~Copet, G.~Synnaeve, and Y.~Adi, ``High fidelity neural audio compression,'' \emph{arXiv preprint arXiv:2210.13438}, 2022.

\bibitem{imagesteg}
F.~Shang, Y.~Lan, J.~Yang, E.~Li, and X.~Kang, ``Robust data hiding for jpeg images with invertible neural network,'' \emph{Neural Networks}, vol. 163, pp. 219--232, 2023.

\bibitem{imagesteg1}
H.~Yang, Y.~Xu, and X.~Liu, ``Dkis: Decay weight invertible image steganography with private key,'' \emph{arXiv preprint arXiv:2311.18243}, 2023.

\bibitem{fang2023flow}
H.~Fang, Y.~Qiu, K.~Chen, J.~Zhang, W.~Zhang, and E.-C. Chang, ``Flow-based robust watermarking with invertible noise layer for black-box distortions,'' in \emph{Proceedings of the AAAI conference on artificial intelligence}, vol.~37, 2023, pp. 5054--5061.

\bibitem{videosteg}
C.~Mou, Y.~Xu, J.~Song, C.~Zhao, B.~Ghanem, and J.~Zhang, ``Large-capacity and flexible video steganography via invertible neural network,'' in \emph{Proceedings of the IEEE/CVF Conference on Computer Vision and Pattern Recognition}, 2023, pp. 22\,606--22\,615.

\bibitem{kingma2018glow}
D.~P. Kingma and P.~Dhariwal, ``Glow: Generative flow with invertible 1x1 convolutions,'' \emph{Advances in neural information processing systems}, vol.~31, 2018.

\bibitem{inn1}
J.~Jing, X.~Deng, M.~Xu, J.~Wang, and Z.~Guan, ``Hinet: Deep image hiding by invertible network,'' in \emph{Proceedings of the IEEE/CVF international conference on computer vision}, 2021, pp. 4733--4742.

\bibitem{inn2}
M.~Xiao, S.~Zheng, C.~Liu, Y.~Wang, D.~He, G.~Ke, J.~Bian, Z.~Lin, and T.-Y. Liu, ``Invertible image rescaling,'' in \emph{Computer Vision--ECCV 2020: 16th European Conference, Glasgow, UK, August 23--28, 2020, Proceedings, Part I 16}.\hskip 1em plus 0.5em minus 0.4em\relax Springer, 2020, pp. 126--144.

\bibitem{panayotov2015librispeech}
V.~Panayotov, G.~Chen, D.~Povey, and S.~Khudanpur, ``Librispeech: an asr corpus based on public domain audio books,'' in \emph{2015 IEEE international conference on acoustics, speech and signal processing (ICASSP)}.\hskip 1em plus 0.5em minus 0.4em\relax IEEE, 2015, pp. 5206--5210.

\bibitem{ardila2019common}
R.~Ardila, M.~Branson, K.~Davis, M.~Henretty, M.~Kohler, J.~Meyer, R.~Morais, L.~Saunders, F.~M. Tyers, and G.~Weber, ``Common voice: A massively-multilingual speech corpus,'' \emph{arXiv preprint arXiv:1912.06670}, 2019.

\bibitem{audioset}
J.~F. Gemmeke, D.~P. Ellis, D.~Freedman, A.~Jansen, W.~Lawrence, R.~C. Moore, M.~Plakal, and M.~Ritter, ``Audio set: An ontology and human-labeled dataset for audio events,'' in \emph{2017 IEEE international conference on acoustics, speech and signal processing (ICASSP)}.\hskip 1em plus 0.5em minus 0.4em\relax IEEE, 2017, pp. 776--780.

\bibitem{defferrard2016fma}
M.~Defferrard, K.~Benzi, P.~Vandergheynst, and X.~Bresson, ``Fma: A dataset for music analysis,'' \emph{arXiv preprint arXiv:1612.01840}, 2016.

\bibitem{pesq}
A.~W. Rix, J.~G. Beerends, M.~P. Hollier, and A.~P. Hekstra, ``Perceptual evaluation of speech quality (pesq)-a new method for speech quality assessment of telephone networks and codecs,'' in \emph{2001 IEEE international conference on acoustics, speech, and signal processing. Proceedings (Cat. No. 01CH37221)}, vol.~2.\hskip 1em plus 0.5em minus 0.4em\relax IEEE, 2001, pp. 749--752.

\end{thebibliography}

\end{document}